# Study of the photo-switching of a Fe(II) chiral complex through linear and nonlinear ultrafast spectroscopy

*Amine Ould Hamouda,[1] Fredéric Dutin,[1] Jérôme Degert,[1] Marc Tondusson,[1] Ahmad Naim,[2] Patrick Rosa,[2] and Eric Freysz*[1]*

[1]Univ. Bordeaux, CNRS, LOMA UMR 5798, 351 cours de la libération 33405 Talence Cedex France.

[2]CNRS, Univ. Bordeaux, Bordeaux INP, ICMCB, UMR 5026, F-33600 Pessac, France.

AUTHOR INFORMATION

**Corresponding Author**

*E-mail:eric.freysz@u-bordeaux.fr

ABSTRACT: Photo-switching the physical properties of molecular systems opens large possibilities for driving materials far from equilibrium toward new states. Moreover, ultra-short pulses of light make it possible to induce and to record photo-switching on a very short time-scale, opening the way to fascinating new functionalities. Among molecular materials, Fe(II) complexes exhibit an ultrafast spin-state transition during which the spin state of the complex switches from a low spin state (LS, S=0) to a high spin state (HS, S=2). The latter process is remarkable: it takes place within ~100 fs with a quantum efficiency of ~100%. Moreover, the spin state switching induces an important shift of the broad metal to ligand absorption band of the complex and it results in large modifications of the physical and chemical properties of the compounds. But, because most of the Fe(II) complexes crystallize in centrosymmetric space groups, this prevents them from exhibiting piezoelectric, ferroelectric as well as second-order nonlinear optical properties such as second harmonic generation (SHG). This considerably limits their potential applications. We have recently synthesized [Fe(phen)$_3$] [$\Delta$-As$_2$(tartrate)$_2$] chiral complexes that crystallize in a noncentrosymmetric 32 space group. Hereafter, upon the excitation of a thin film of these complexes by a femtosecond laser pulse and performing simultaneously transient absorption (TRA) and time-resolved SHG (TRSH) measurements, we have recorded the ultrafast LS to HS switching. Whereas a single TRA measurement only gives partial information, we demonstrate that TRSH readily reveals the different mechanisms in play during the HS to LS state relaxation. Moreover, a simple model makes it possible to evaluate the relaxation times as well as the hyperpolarizabilities of the different excited states through which the system travels during the spin-state transition.

**TOC GRAPHICS**

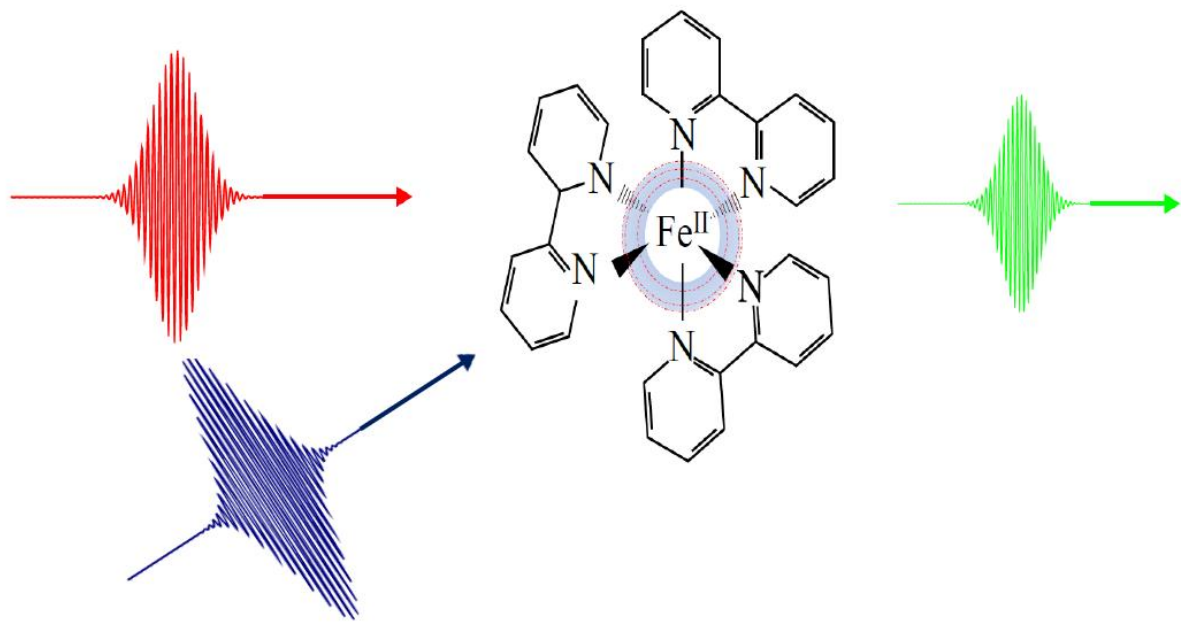

Researches devoted to materials with dedicated properties that can be easily synthesized on a large scale have always attracted much attention. More recently, the possibility to photo-switch their properties far from equilibrium toward new states has further expanded the possibilities of applications of such materials. Along this line, organic materials are of particular interest beacause they can be easily synthesized by "wet" chemistry using a large variety of organic molecules.[1] Because the presence of metal to ligand charge transfer (MLCT) or ligand to metal charge transfer (LMCT) transitions, organo-metallic and coordination metal complexes represent an original and growing class of molecular materials.[2] These complexes offer many prospects for magnetic, electric or chemical sensors.[3–6] Moreover, these materials enable low-cost production on various substrates.[7,8] The addition of switchability has further triggered interest in their applications in very different technologies.[9,10] Photo-switching of the linear optical properties of these complexes can be induced by many means.[11–16] The switching can be ultrafast if one photo-switches the ligand or the spin state of the metallic ion. The spin state photo-switching of Fe(II) ions from low spin (LS) S=0 to high spin (HS) S=2 is of particular interest since it is remarkably fast (~100 fs) and exhibits a quantum efficiency near unity.[17–24] Moreover, because an important list of proteins have Fe(II) centers in their active sites, the study of the photo-switching of iron(II) is also of crucial importance in life sciences.[25–28] The mechanism giving rise to this ultrafast switching in Fe(II) complexes is still the subject of debate. To address this issue, ultrafast photo-switching of the spin state of some Fe(II) complexes has been recorded by performing transient absorption measurement or transient X-ray diffraction both in solution and for bulk materials.[17–24,29–31] As we already mentioned, the photo-switching naturally results in a change of the linear optical properties of complexes. But it should also impact nonlinear optical properties. However, to the best of our knowledge, no experiment has been reported so far on ultrafast photo-switching of the nonlinear optical (NLO) properties of coordination complexes. In this context, it should be stressed that most complexes usually crystallize in centrosymmetric space groups, which prevents them from exhibiting properties such as piezoelectricity, ferroelectricity, as well as second-order optical properties such as electro-optical effect, second-harmonic (SH) as well as sum-frequency generation. The absence of these properties limits considerably their potential application in nonlinear optics. We recently synthesized chiral compounds based on [Fe(phen)$_3$]$^{2+}$ complexes that are stabilized by adding different chiral counter-ions.[32] The addition of chiral counter-ions prevents them crystallizing in centrosymmetric space groups. Consequently, we showed that these crystals, belonging to the 32 symmetry class, exhibit a sensible second order optical coefficient $\chi^{(2)}_{111} = 6.5 \pm 0.5\ 10^{-12}\ m.V^{-1}$.[33] The noticeable MLCT absorption band of these crystals, centered at 500 nm, indicates Fe(II) is in the LS state and the crystals remain in this spin state at high temperatures. Since we previously showed that, for [Fe(phen)$_3$]$^{2+}$ in solution ultrafast photo-switching of the spin state can be recorded,[34,35] one could wonder if the same photo-switching could be recorded by exciting those noncentrosymmetric crystals with femtosecond pulses. We demonstrate hereafter that this is indeed the case. We show that the spin state photo-switching results in a modulation of the second-order NLO properties of these crystals. We also show that time-resolved second-harmonic (TRSH) experiments make it possible to readily record and reveal the kinetic of the the ultrafast photo-switching of Fe(II) spin state from the LS state to the HS state. We also propose a simple model that accounts for our experimental data. Finally, we demonstrate that the TRSH experiment along with conventional time-resolved absorption (TRA) measurements make it possible to evaluate the second-order nonlinear optical properties of the excited state involved in the spin state photo-switching.

**Experimental results:**

The first TRA experiment was performed dissolving [Fe(2-CH$_3$-phen)$_3$](BF$_4$)$_2$ complexes in acetonitrile. To minimize the impact of the pump pulse on the solution, the experiment was performed in a 1 mm thick quartz flow-through cell. The central wavelength of the pump pulse was fixed to be ~480 nm, close to the center of the broad metal to ligand charge transfer (MLCT) absorption band in the LS state of the Fe(II) complex. The central wavelength of the probe pulse was chosen centered about 540 nm which is also within the broad absorption band of the sample in the LS state of Fe(II).[36] In agreement with our previous study, Fig. 1 indicates that upon excitation, the absorption of the sample rapidly decreases and then slowly recovers its initial state with a time constant $\tau_R$ ~1.0±0.1 ns.

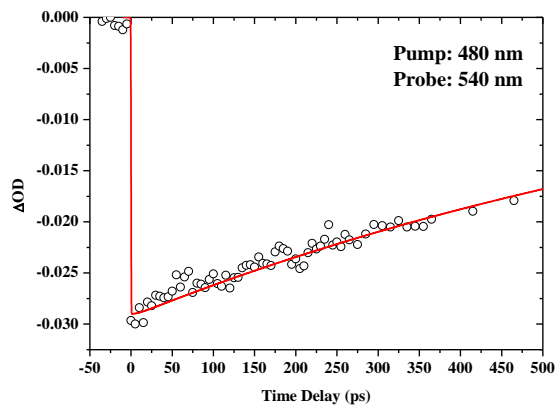

*Figure 1: Time resolved absorption experiment for [Fe(2-CH$_3$-phen)$_3$] in acetonitrile. The pump and probe central wavelengths are fixed at 480 and 540 nm, respectively. Dots are experimental data, and the solid red line corresponds to the fit of the data.*

We performed the same experiment on crystallites of [Fe(phen)$_3$]($\Delta$-As$_2$(tartrate)$_2$) by drop-casting a droplet of a saturated solution on a microscope cover glass. We showed previously that the as-prepared crystallites make it possible to record SH.[33] The evolution of the TRA measurement performed on this sample is presented in Fig. 2a. Upon excitation, one can notice that the optical density of the sample rapidly decreases, then reaches a small plateau which relaxes in ~50±5 ps, and finally slowly relaxes back towards its initial state with a constant time $\tau_R$ ~0.9±0.1 ns. We also measured the evolution of the maximal change of the optical density versus the pump intensity and found it was linear (Fig. 2b). Let us mention that, when excited by a 6 µJ pump pulse, the induced change of the optical density of our sample is 3.0±0.3%.

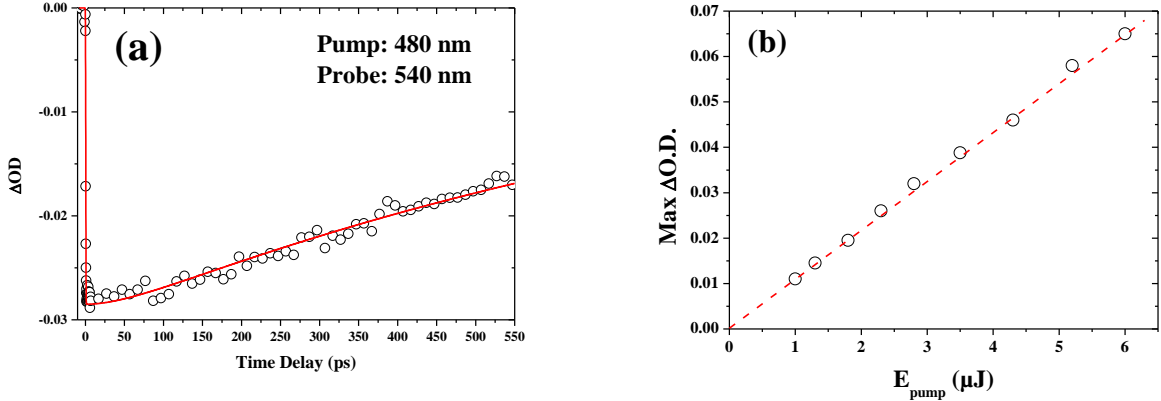

**Figure 2** (*a*) *Time resolved absorption experiment for bulk [Fe(phen)$_3$]($\Delta$-As$_2$(tartrate)$_2$). The pump and probe central wavelengths are fixed at 480 and 540 nm, respectively. Dots and the solid line are experimental data and the fit of these data, respectively. (b) Evolution of the maximal change of the optical density versus the pump pulse energy. Dots and the dashed line are experimental data and the linear fit of these data, respectively.*

We then performed the TRSH measurements. The central wavelength of the pump pulse was kept at 480 nm and the central wavelength of the probe pulse was fixed at 1060 nm. The evolution of the SH signal upon excitation is presented in Fig. 3. One can notice that upon excitation the SH signal rapidly decreases. Just after excitation it rapidly recovers 80 % and 91 % of its initial value after ~5 ps and ~50 ps, respectively and then more slowly relaxes. We have also recorded the evolution of the amplitude of the TRSH signal at its minimum versus the pump and probe intensities. While the temporal shape remains similar, the amplitude of the signal was found to evolve linearly with respect to the pump intensity (Fig. 4a). Similarly, we found that the amplitude of the SH signal evolves like the square of the probe intensity (Fig. 4b). When the energy of the pump pulse was fixed at 6 μJ, the amplitude change of SH signal, hereafter labeled

$R(2\omega) = [I_{pump\_on}(2\omega) - I_{pump\_off}(2\omega)]/I_{pump\_off}(2\omega)$, was found to be $R(2\omega) = 5.0 \pm 0.5$ %.

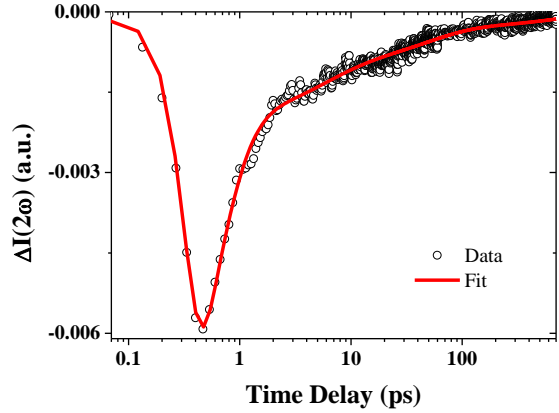

*Figure 3*: *Time-resolved second-harmonic generation experiment for $[Fe(phen)_3](\Delta\text{-}As_2(tartrate)_2)$. The pump and probe central wavelengths are fixed at 480 and 1060 nm, respectively. The dots and the solid line are the experimental data and the fit of these data, respectively.*

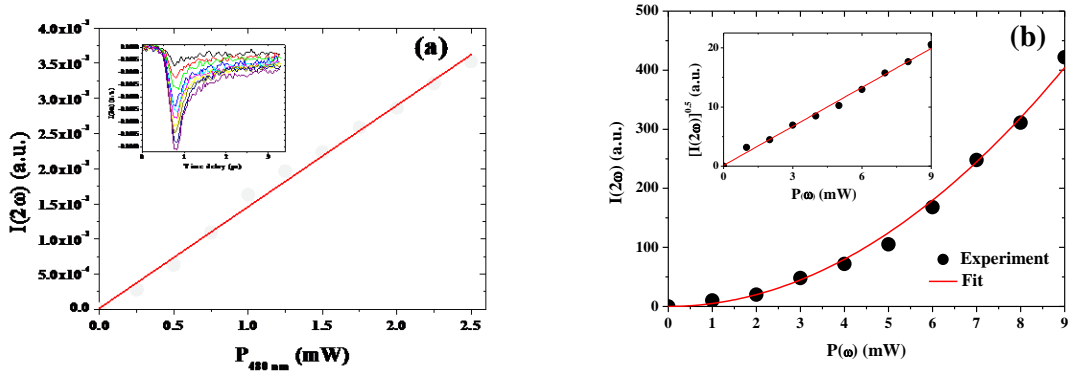

*Figure 4:* *(a) Evolution of the second-harmonic signal at its minimum versus the pump intensity. Dots and the solid line are experimental data and a linear fit, respectively. The inset presents the evolution of the SH signal versus the pump-probe delay for different pump intensities. (b) Evolution of the amplitude of the second-harmonic signal at its minimum versus the probe intensity. The inset presents the evolution of the square root of this signal. Dots and solid lines are respectively experimental data and quadratic fits of these data.*

## **Model for TRSH experiment**

To model the TRSH measurements, we considered that the intensity of the SH signal generated by the probe pulse writes:

$$I(2\omega) \sim \left[ L\, N_T \langle \beta^{(2)}(2\omega,\omega) \rangle I(\omega) \right]^2, \quad (1)$$

where L, $N_T$, $\langle \beta^{(2)}(2\omega,\omega) \rangle$ and I(ω) are the local field factor, number of chiral Fe(II) complexes per unit volume, the hyperpolarizability of the complexes at frequency ω, and the intensity of the probe pulse, respectively[37] and $\langle \rangle$ denotes an angular average with respect to the randomly oriented crystallites of the sample.[38] Hereafter, we will denote $\beta'^{(2)} = L \langle \beta^{(2)}(2\omega,\omega) \rangle$ the effective hyperpolarizability corresponding to the averaged hyperpolarizability corrected from the local field factor L. To account for the evolution of the SH signal upon excitation, we have to consider that the second order optical hyperpolarizability of the complexes depends on the excited state in which the Fe(II) complexes are set upon excitation. Accordingly, the SH signal upon excitation of the sample writes:

$$I(2\omega, \tau) \sim \left[ \sum_{i=0}^{j} N_i(\tau) \beta'^{(2)}_i I(\omega) \right]^2, \quad (2)$$

where τ, $N_i(\tau)$ and $\beta'^{(2)}_i$ are the time delay between the pump and the probe pulse, the population, and the effective hyperpolarizability of the i[th] excited state, respectively. Prior to its excitation, the sample is suppose to be in its fundamental state i=0. One should notice that $N_T = \sum_{i=0}^{j} N_i(\tau)$. Thus, prior to excitation, we have $N_i = 0$ for i>0 and $N_{i=0} = N_T$ and Eq.(1) writes $I(2\omega) \sim [ N_T \beta'^{(2)}_0 I(\omega) ]^2$. Accordingly, one can rewrite Eq. (2) as follow:

$$I(2\omega, \tau) \sim \left[ N_T \beta'^{(2)}_0 - \sum_{i=1}^{j} N_i(\tau) \Delta\beta'^{(2)}_{0i} \right]^2 I^2(\omega), \quad (3)$$

where $\Delta\beta'^{(2)}_{0i} = \beta'^{(2)}_0 - \beta'^{(2)}_i$. If, as observed in our experiment, the change in the SH signal is small, then a Taylor expansion of Eq. (3) yields:

$$I(2\omega, \tau) \sim \left[ N_T \beta'^{(2)}_0 I(\omega) \right]^2 - 2 N_T \beta'^{(2)}_0 I^2(\omega) \sum_{i=1}^{j} N_i(\tau) \Delta\beta'^{(2)}_{0i}$$

Our experimental set-up records the difference in the SH signal with and without the pump beam, hence, the TRSH signal writes:

$$S_{TRSH}(2\omega,\tau) \sim -2\,N_T \beta_0'^{(2)} I^2(\omega) \sum_{i=1}^{j} N_i(\tau)\,\Delta\beta_{0i}'^{(2)}. \quad (4)$$

In agreement with our experimental results, the latter expression stresses that the TRSH signal evolves linearly with respect to the population of the excited states $N_i(\tau)$ and quadratically with respect to the probe intensity. Furthermore, for a given probe intensity, the ratio of Eq. (4) and (1) writes

$$R(2\omega,\tau) = -2 \sum_{i=1}^{j} P_i(\tau)\, \frac{\Delta\beta_{0i}'^{(2)}}{\beta_0'^{(2)}}, \quad (5)$$

where $P_i(\tau)=N_i(\tau)/N_T$ is the fractional population of the $i^{th}$ excited state at the delay $\tau$. Hence, to quantify the relative change of the hyperpolarizability of the Fe(II) complexes in their excited states, $\Delta\beta_{0i}'^{(2)}/\beta_0'^{(2)}$, one needs to evaluate $P_i(\tau)$. In the TRA experiment, within the Beer-Lambert's regime and along the MLCT absorption band, the change of the optical density of the sample is directly linked to the variation of the population of the complexes in their excited states. Hence running both a TRA and TRSH measurements should make it possible to evaluate the population of the exited states, their relaxation constant as well as the variation of hyperpolarizability of the complexes in their excited states.

**Analysis of experimental data.**

*a. TRA experiment of the complexes in solution and in the bulk*

It has been previously shown that upon excitation of the MLCT absorption band the spin state of the Fe(II) complexes switches very rapidly from S=0 to S=2 with a high quantum yield near unity.[17–24] Once in their excited high spin state and within a few picoseconds, the complexes relax towards a non-vibrationally excited HS state. Afterwards, and on a longer time scale, they relax back towards their fundamental LS state. All these relaxation processes are non-radiating and the relaxation scheme accepted so far is depicted in Fig. 5a.

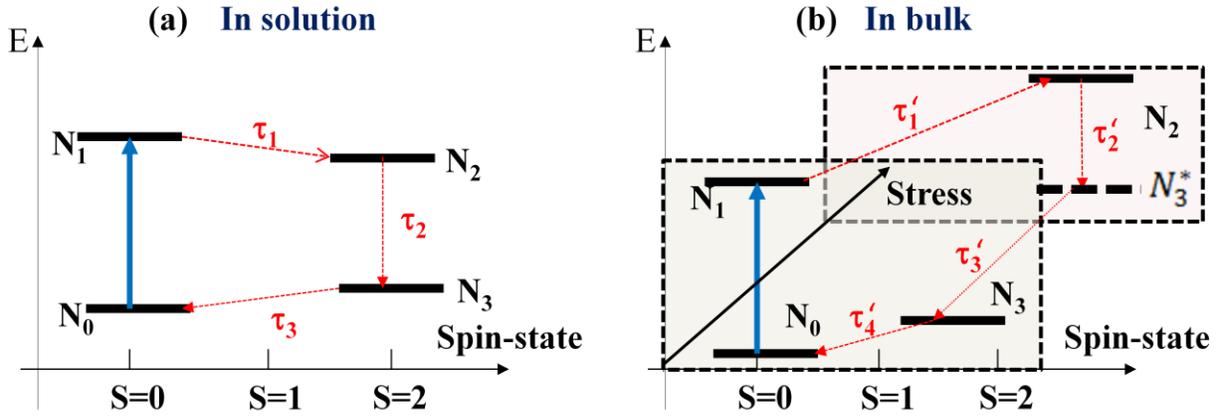

*Figure 5: Schematic representation of photo-switching pathway for complexes a) in solution and b) in the bulk state.*

Accordingly, we have fitted the experimental data by optimizing through a $\chi^{(2)}$ minimizing process a set of initial parameters used to numerically integrate the following set of differential equations for the fractional populations $P_i=N_i/N_T$ of the different excited states (solid lines in Fig. 1 and Fig. 2).

$$\frac{dP_1}{dt} = \frac{\sigma_p I_{p0}}{\sqrt{\pi}\tau_p} \exp\left(-\frac{(t-t_0)^2}{\tau_p^2}\right) - \frac{P_1}{\tau_1},$$

$$\frac{dP_2}{dt} = \frac{P_1}{\tau_1} - \frac{P_2}{\tau_2},$$

$$\frac{dP_3}{dt} = \frac{P_2}{\tau_2} - \frac{P_3}{\tau_3},$$

where $I_{p0}$, $t_0$ and $\tau_p$ are the intensity, the peak position of a Gaussian shaped pulse and the duration of the pump pulse, respectively. This set of partial differential equations is supplemented with the following boundary conditions: for t=0, $P_0$=1, $P_1$=$P_2$=$P_3$=0 with $P_0$+$P_1$+$P_2$+$P_3$=1. As already mentioned, for the probe wavelength used to record our data, the TRA is mainly sensitive to the evolution of the population of the complexes in their fundamental state. We were able to fit the TRA data presented in Fig. 1 considering $\tau_p=\tau_1=\tau_2=400\pm50$ fs and $\tau_3=1.0\pm0.1$ ns. The latter is related to the relaxation of the vibrationally cooled high spin state.[36] Note that $\tau_p$ is larger than the actual pump pulse duration ($\tau_{pu}\sim80$ fs). This discrepancy is due to the so-called "windowing" effect which is induced by the small angle between the two weakly focused pump and probe pulses. Accordingly, hereafter, we will consider that our instrumental response function limited by this effect is ~400 fs. It is also worth mentioning that in order to retrieve the contribution of the two other time constants (i.e. $\tau_1\sim300$ fs and $\tau_2\sim8.6$ ps) we have shown in a previous work, we had to perform the TRA experiment keeping the same excitation wavelength but setting the central wavelength of the probe pulse around absorption bands of the excited states.[35] To improve the fit of the data presented in Fig. 2, besides $\tau_p=\tau_1=400\pm50$ fs, we had to consider two time constants $\tau_2=50\pm5$ ps and $\tau_3=900\pm100$ ps. The optimum fit of the data with and without the introduction of the constant $\tau_2$ is displayed in the supplementary material S1. Note the time constant $\tau_2$ is larger than $\tau_2$ we found for the complexes in solution.

### b. TRSH experiments of the complexes in bulk

According to Eq. (4) and the relaxation scheme of the complexes in solution (Fig. 5a), the temporal evolution of the TRSH signal depends on the evolution of the population $N_1$, $N_2$ and $N_3$ in the different excited states. Hence, one should expect to be able to fit the experimental data using the relaxation time constant hereafter labelled $\tau'_i$ determined to fit the TRA experiment. In fact, to account for the abrupt rapid drop of the SH signal just after excitation of the sample, besides the three time constants $\tau'_1=400\pm20$ fs, $\tau'_3=45\pm3$ ps and $\tau'_4=800\pm100$ ps, we had to introduce another relaxation time constant $\tau'_2=5.0\pm0.4$ ps. The impact of the introduction of the different sets of time constants in the fit of the experimental data is presented in the supplementary material S3. One will clearly notice that at least three time constants are required to properly fit our experimental data. The introduction of the relaxation time constant $\tau_2'$ makes it possible to recover more closely the evolution of the TRSHG signal at the early stage of the relaxation. The value of the time constant $\tau_2'$ is close to the value found in solution for vibrational cooling of the complexes in the HS state.[24] Therefore, it is likely associated to the relaxation of the vibrationally excited state with a spin state S=2. Hence, to account for the constant time $\tau_2'$, one has to add a new excited state labeled hereafter $N_3^*$ whose energy level is in between the state $N_2$ and $N_3$. It is the latter excited state which relaxes with the constant $\tau_3'$=45±3 ps. As a consequence, the relaxation process for the complexes in the bulk state is depicted by Fig. 5b instead of Fig. 5a. It means that, neglecting the contribution of any other relaxation process, one has to supplement our sequential relaxation of the fractional population with an additional differential equation:

$$\frac{dP_1}{dt} = \frac{\sigma_p I_{p0}}{\sqrt{\pi}\tau_p} \exp\left(-\frac{(t-t_0)^2}{\tau_p^2}\right) - \frac{P_1}{\tau_1'},$$

$$\frac{dP_2}{dt} = \frac{P_1}{\tau_1'} - \frac{P_2}{\tau_2'},$$

$$\frac{dP_3^*}{dt} = \frac{P_2}{\tau_2'} - \frac{P_3^*}{\tau_3'}, \qquad (6)$$

$$\frac{dP_3}{dt} = \frac{P_3^*}{\tau_3'} - \frac{P_3}{\tau_4'},$$

Accordingly, adjusting the parameters in front of the population of the excited states, and using this set of time constants $\tau_1'$, $\tau_2'$, $\tau_3'$ and $\tau_4'$ partly determined by the TRA experiment, we were able to nicely reproduce of TRSH data (Fig 3, solid red line). The fitted amplitude $A_i$ of the pre-exponential factor associated to the different relaxation time constant as well as the uncertainty on the latter parameters are displayed in table 1. The pre-exponential factors determined by the fitting procedure are proportional to the population and the hyperpolarizability of the excited state involved in the spin-state transition. Noticeably in our experience, the great difference in the relaxation time $\tau'_1 \ll \tau'_2 \ll \tau'_3 \ll \tau'_4$ makes it possible to consider that, for some given delay T, all the population of complexes are within a given excited state.

| | |
|---|---|
| $A_1 = (8.5 \pm 0.3)*10^{-4}$ | $\tau'_1 = 0.4 \pm 0.02$ ps |
| $A_2 = (2.0 \pm 0.1)*10^{-4}$ | $\tau'_2 = 5.0 \pm 0.4$ ps |
| $A_3 = (1.0 \pm 0.05)*10^{-4}$ | $\tau'_3 = 45 \pm 3$ ps |
| $A_4 = (3.0 \pm 0.3)*10^{-5}$ | $\tau'_4 = 800 \pm 100$ ps |

*Table 1: Parameters obtained by fitting the relaxation of the relaxation of the time resolved second harmonic generation through the numerical integration of Eq. (6).*

Hence, according to Eq. (5), knowing the relative variation of the SH signal R(2ω,□), the relative variation of the population within the excited state P$_i$(t), one is able to evaluate the hyperpolarizability $\beta_i'^{(2)}$ of the different excited states. According to Eq. (5) the later writes:

$$\beta_i'^{(2)} = \beta_0'^{(2)} \left(1 - \frac{R_i(2\omega)}{2P_i}\right),$$

where R$_i$(2ω) is the relative variation of the SHG signal of the i$^{th}$ excited state. Since, for a given pump intensity, thanks to the very high quantum yields of the spin state transition in Fe(II) complexes which, hereafter, is considered to be ~100%, the TRA and TRSH experiments make it possible to evaluate P$_i$ and R$_i$(2ω), one is then able to evaluate the ratio $\beta_i'^{(2)}/\beta_0'^{(2)}$. Since we already estimated $\chi_{111}^{(2)} = \chi_0^{(2)} \sim N_T \beta_0'^{(2)}$,[33] we can evaluate $\chi_{N_i}^{(2)}$ the optical susceptibility of the crystal in the different excited states ($\chi_{N_i}^{(2)} = \chi_{111}^{(2)*} \sim N_T \beta_i'^{(2)}$). The results of our computations are detailed in Table 2.

| | |
|---|---|
| $\chi_0^{(2)} = \chi_{N_0}^{(2)}$ | $6.5 \pm 0.5\ 10^{-12}$ m.V$^{-1}$ |
| $\chi_{N_1}^{(2)}$ | $1.0 \pm 0.5\ 10^{-12}$ m.V$^{-1}$ |
| $\chi_{N_2}^{(2)}$ | $2.0 \pm 0.5\ 10^{-12}$ m.V$^{-1}$ |
| $\chi_{N_3}^{(2)} \cong \chi_{N_3^*}^{(2)}$ | $6.0 \pm 0.5\ 10^{-12}$ m.V$^{-1}$ |

**Table 12:** *Second-order optical susceptibilities of the Fe(II) complexes in their different excited states.*

## **Discussion**

The large susceptibility of our crystalline sample in its fundamental state is mainly related to a resonant enhancement of the second-order optical susceptibility of this compound at 540 nm which is associated to the MLCT absorption band.[33] When the complexes are brought in the MLCT excited states, the second-order susceptibility of the micro-crystallites drops abruptly to about 1 pm.V$^{-1}$. The latter value is about the value we measured for this sample when the excitation is out of resonance with the MLCT absorption band.[33] Once in the high spin state (S=2), the second-order optical susceptibility of the complex increases to ~2 pm.V$^{-1}$ and ~6 pm.V$^{-1}$ in its vibrationally and non-vibrationally excited state, respectively. These large values likely indicate that, once in their HS state, the hyperpolarizability of the complexes is also resonantly enhancement due to the presence of a $^5T_2$ to $^5E$ or $^5$MLCT absorption bands.[36]

While most of the time constant associated to the relaxation of the excited LS state, vibrationally excited and vibrationally cooled HS state we used to fit our TRSH and TRA agree well with the literature, one may question about the origin of the state N$^*_3$ whose time constant is $\tau_3$~50 ps. It should be noted that the latter cannot be associated with vibrational cooling of the complexes which usually lasts less than 10 ps. To account for this state, we propose the following process. Upon LS to HS switching, there is a change of volume of the complexes associated to the change in the Fe-N bond distance. This change of volume induces a local stress within the crystallites. Hence, as cartooned in Fig.5b we propose that $N^*_3$ is related to a vibrationnally cooled high spin state in a locally stressed environment. The latter relaxes with a time constant $\tau_3$ associated with the release of the photo-induced stress through the propagation of an acoustic wave at the speed of sound $v_s$ in the micro crystallites of characteristic size L. A rough estimate of this constant time $\tau_3 \sim L/2v_s \sim 41$ to 150 ps where $v_s \sim 3$ to $6\ 10^3$ m.s$^{-1}$ and L~0.9 to 0.5 µm is in good agreement with our data. As suggested by one of our reviewer, it should be noted that this photo-induced stress probably also affects the complexes remaining in the LS state in the close neighborhood of the excited complexes. Hence, the hyperpolarizability of both photo-excited complexes and the neighboring complexes remaining in the LS should be affected. The impact of the latter can be accounted considering that upon excitation the local field factor L (see Eq. (1)) is modified. Note the picture we just have drawn is based on the assumption we made which consists in considering that no other mechanism triggered by the pump pulse, and described by the set of differential equations we used, is taking place. Further experiments performed probing both TRA and TRSH at other wavelengths will help to confirm this assumption.

**Conclusions and prospects**

We have performed ultrafast photo-switching of the nonlinear optical properties of chiral SCO materials. The modulation of the second-order optical nonlinear properties takes place on an ultrashort time scale. We have shown that, upon excitation, the amplitude of the modulation is large. Moreover, we have shown that the material we study recovers its initial properties on a relatively short time scale. In fact, this study opens interesting possibilities we would like to explore in near future such as studying the photo-switching properties of few layers of SCO compounds or the modification of the circular dichroism of these compounds upon excitation. The TRSH technique we have proposed should also make it possible to reveal and follow modification of the crystalline structure upon excitation in different SCO materials as well as many others such as cyanide-bridged bimetallic assemblies or insulating-metal charge-transfer systems and photochromic materials. Finally, it is worth mentioning that upon excitation all the nonlinear optical properties of the material, such as the third-order optical nonlinear optical properties, should be also affected, works along this line being in progress in our university.

**Experimental methods**

   a. **Sample preparation**

Small crystallites of [Fe(2-CH$_3$-phen)$_3$](BF$_4$)$_2$ (2-CH$_3$-phen =2-methyl-1,10-phenanthroline) complexes were dissolved in acetonitrile (0.1 mg/ml). The synthesis of enantiomeric crystals of [Fe(phen)$_3$](As$_2$(tartrate)$_2$) was reported previously.[32] The sample we studied was prepared dissolving in N,N'-Dimethylformamide micro-crystallites of the compounds (10–15 mg/mL). Droplets of the solution ($\approx$4×10 uL) were drop-cast on a microscope cover glass and dried under rough vacuum for a couple hours until perfectly dry. The as prepared dried thin film comprises small crystallites likely sub-micrometric in size randomly dispersed over the cover glass.

   b. **Optical set-up:**

The experimental set-up is depicted in Fig. 6. The samples are excited and probed by pulses yielded by two optical parametric amplifiers (OPA, TOPAS, Ligth conversion) which are pumped by 1 mJ pulses of $\tau_{laser}$~50 fs duration (full width at half maximum: FWHM), centered at 800 nm. They are delivered at a repetition rate of 1 kHz by a Ti:Sapphire regenerative amplifier (Coherent, Legend). The central wavelength of the pump (pu) and probe (pr) pulses are tunable in between 1020 nm and 2800 nm. Within this spectral range, the pump and probe pulse duration ( $\tau_{pu}$ and $\tau_{pr}$, respectively) is about 120 fs (FWHM). The pump and probe pulses are slightly out of focus on the same spot and at different angles of incidence onto the samples. A delay line makes it possible to delay in time the pump and the probe pulses. With this set-up, we performed either time resolved absorption (TRA) or time resolved second harmonic (TRSH) measurements. For TRA, the intensity of the pump pulse (output of OPA 1) was modulated by a mechanical chopper (Thorlabs, MC2000B) and we measured the evolution of the intensity of the probe pulse (second harmonic of OPA 2) transmitted by the sample versus the delay τ between the pump and the probe pulses. The intensity of the probe pulse was recorded by a photodiode (Phd Ref$_1$: Thorlabs DET36A2) whose output was connected to a lock-in amplifier (Stanford Research Instrument SR830) synchronized with the mechanical chopper. For TRSH, the SH signal generated within the sample was sent through a dichroic filter which transmitted the SH signal and blocked the fundamental probe spectrum (output of OPA 2). The transmitted SH signal was focused on the entrance slit of a monochromator whose output was detected by a photomultiplier tube (PMT). The latter electric signal was injected in the lock-in amplifier whose output was recorded versus pump-probe time delay τ.

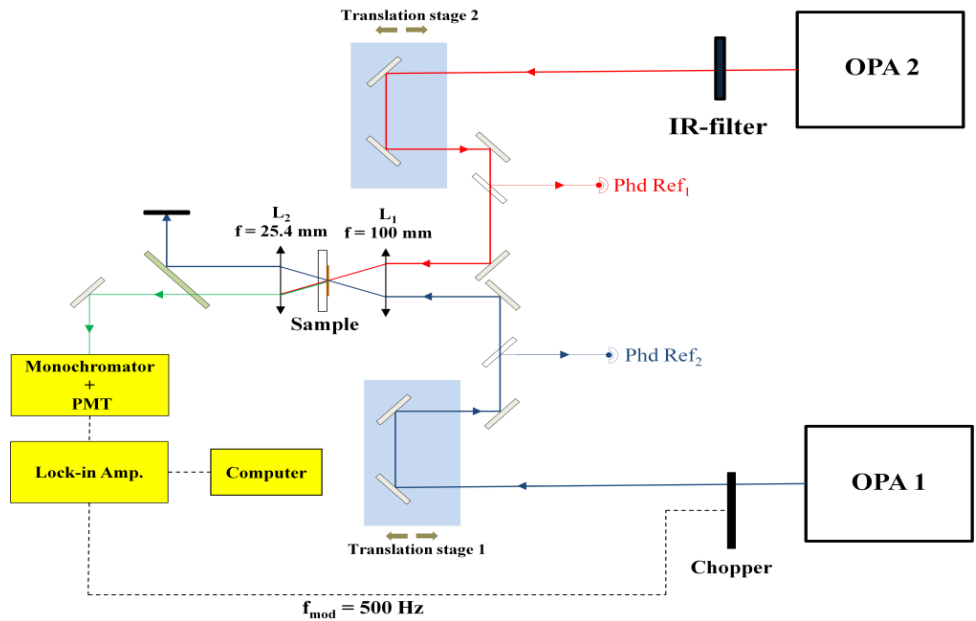

*Figure 6*: *Optical setup in TRSH configuration: OPA: optical parametric amplifier. $L_i$: lenses. PMT: photomultiplier tube. Lock-in Amp.: lock-in amplifier. Phd: photodiode*

AUTHOR INFORMATION

The authors declare no competing financial interests.

ACKNOWLEDGMENT

The Conseil Régional Nouvelle Aquitaine and FEDER are thanked for funding the equipments of the COLA platform at LOMA.

REFERENCES

(1) Jones, W. *Organic Molecular Solids Properties and Applications*; CRC Press: Boca Raton, FL, 1997.
(2) Ashcroft, C. M.; Cole, J. M. Molecular Engineering of Organic and Organometallic Second-Order Nonlinear Optical Materials. In *Handbook of Organic Materials for Electronic and Photonic Devices*; Elsevier, 2019; pp 139–176. https://doi.org/10.1016/B978-0-08-102284-9.00005-X.
(3) Prins, F.; Barreiro, A.; Ruitenberg, J. W.; Seldenthuis, J. S.; Aliaga-Alcalde, N.; Vandersypen, L. M. K.; van der Zant, H. S. J. Room-Temperature Gating of Molecular Junctions Using Few-Layer Graphene Nanogap Electrodes. *Nano Lett.* **2011**, *11* (11), 4607–4611. https://doi.org/10.1021/nl202065x.
(4) Dugay, J.; Giménez-Marqués, M.; Venstra, W. J.; Torres-Cavanillas, R.; Sheombarsing, U. N.; Manca, N.; Coronado, E.; van der Zant, H. S. J. Sensing of the Molecular Spin in Spin-Crossover Nanoparticles with Micromechanical Resonators. *J. Phys. Chem. C* **2019**, *123* (11), 6778–6786. https://doi.org/10.1021/acs.jpcc.8b10096.
(5) Dugay, J.; Giménez-Marqués, M.; Kozlova, T.; Zandbergen, H. W.; Coronado, E.; van der Zant, H. S. J. Spin Switching in Electronic Devices Based on 2D Assemblies of Spin-Crossover Nanoparticles. *Adv. Mater.* **2015**, *27* (7), 1288–1293. https://doi.org/10.1002/adma.201404441.
(6) Coronado, E.; Giménez-Marqués, M.; Espallargas, G. M.; Brammer, L. Tuning the Magneto-Structural Properties of Non-Porous Coordination Polymers by HCl Chemisorption. *Nat. Commun.* **2012**, *3* (1). https://doi.org/10.1038/ncomms1827.
(7) Bousseksou, A.; Molnár, G.; Salmon, L.; Nicolazzi, W. Molecular Spin Crossover Phenomenon: Recent Achievements and Prospects. *Chem. Soc. Rev.* **2011**, *40* (6), 3313. https://doi.org/10.1039/c1cs15042a.
(8) Cavallini, M. Status and Perspectives in Thin Films and Patterning of Spin Crossover Compounds. *Phys. Chem. Chem. Phys.* **2012**, *14* (34), 11867. https://doi.org/10.1039/c2cp40879a.
(9) Shepherd, H. J.; Gural'skiy, I. A.; Quintero, C. M.; Tricard, S.; Salmon, L.; Molnár, G.; Bousseksou, A. Molecular Actuators Driven by Cooperative Spin-State Switching. *Nat. Commun.* **2013**, *4*. https://doi.org/10.1038/ncomms3607.
(10) Salmon, L.; Molnár, G.; Zitouni, D.; Quintero, C.; Bergaud, C.; Micheau, J.-C.; Bousseksou, A. A Novel Approach for Fluorescent Thermometry and Thermal Imaging Purposes Using Spin Crossover Nanoparticles. *J. Mater. Chem.* **2010**, *20* (26), 5499. https://doi.org/10.1039/c0jm00631a.
(11) Gütlich, P.; Hauser, A. Thermal and Light-Induced Spin Crossover in Iron(II) Complexes. *Coord. Chem. Rev.* **1990**, *97*, 1–22. https://doi.org/10.1016/0010-8545(90)80076-6.
(12) Bartual-Murgui, C.; Akou, A.; Thibault, C.; Molnár, G.; Vieu, C.; Salmon, L.; Bousseksou, A. Spin-Crossover Metal–organic Frameworks: Promising Materials for Designing Gas Sensors. *J Mater Chem C* **2015**, *3* (6), 1277–1285. https://doi.org/10.1039/C4TC02441A.
(13) Gutlich, P.; Ksenofontov, V.; Gaspar, A. Pressure Effect Studies on Spin Crossover Systems. *Coord. Chem. Rev.* **2005**, *249* (17–18), 1811–1829. https://doi.org/10.1016/j.ccr.2005.01.022.

(14) Gütlich, P. G. Photoswitchable Coordination Compounds. *Coord. Chem. Rev.* **2001**, *219–221*, 839–879. https://doi.org/10.1016/S0010-8545(01)00381-2.
(15) Southon, P. D.; Liu, L.; Fellows, E. A.; Price, D. J.; Halder, G. J.; Chapman, K. W.; Moubaraki, B.; Murray, K. S.; Létard, J.-F.; Kepert, C. J. Dynamic Interplay between Spin-Crossover and Host−Guest Function in a Nanoporous Metal−Organic Framework Material. *J. Am. Chem. Soc.* **2009**, *131* (31), 10998–11009. https://doi.org/10.1021/ja902187d.
(16) Bousseksou, A.; Negre, N.; Goiran, M.; Salmon, L.; Tuchagues, J.-P.; Boillot, M.-L.; Boukheddaden, K.; Varret, F. Dynamic Triggering of a Spin-Transition by a Pulsed Magnetic Field. *Eur. Phys. J. B-Condens. Matter Complex Syst.* **2000**, *13* (3), 451–456.
(17) Huse, N.; Cho, H.; Hong, K.; Jamula, L.; de Groot, F. M. F.; Kim, T. K.; McCusker, J. K.; Schoenlein, R. W. Femtosecond Soft X-Ray Spectroscopy of Solvated Transition-Metal Complexes: Deciphering the Interplay of Electronic and Structural Dynamics. *J. Phys. Chem. Lett.* **2011**, *2* (8), 880–884. https://doi.org/10.1021/jz200168m.
(18) Bertoni, R.; Lorenc, M.; Tissot, A.; Servol, M.; Boillot, M.-L.; Collet, E. Femtosecond Spin-State Photoswitching of Molecular Nanocrystals Evidenced by Optical Spectroscopy. *Angew. Chem. Int. Ed.* **2012**, *51* (30), 7485–7489. https://doi.org/10.1002/anie.201202215.
(19) Smeigh, A. L.; Creelman, M.; Mathies, R. A.; McCusker, J. K. Femtosecond Time-Resolved Optical and Raman Spectroscopy of Photoinduced Spin Crossover: Temporal Resolution of Low-to-High Spin Optical Switching. *J. Am. Chem. Soc.* **2008**, *130* (43), 14105–14107. https://doi.org/10.1021/ja805949s.
(20) Bressler, C.; Milne, C.; Pham, V.-T.; ElNahhas, A.; van der Veen, R. M.; Gawelda, W.; Johnson, S.; Beaud, P.; Grolimund, D.; Kaiser, M.; et al. Femtosecond XANES Study of the Light-Induced Spin Crossover Dynamics in an Iron(II) Complex. *Science* **2009**, *323* (5913), 489–492. https://doi.org/10.1126/science.1165733.
(21) Huse, N.; Kim, T. K.; Jamula, L.; McCusker, J. K.; de Groot, F. M. F.; Schoenlein, R. W. Photo-Induced Spin-State Conversion in Solvated Transition Metal Complexes Probed via Time-Resolved Soft X-Ray Spectroscopy. *J. Am. Chem. Soc.* **2010**, *132* (19), 6809–6816. https://doi.org/10.1021/ja101381a.
(22) Cammarata, M.; Bertoni, R.; Lorenc, M.; Cailleau, H.; Di Matteo, S.; Mauriac, C.; Matar, S. F.; Lemke, H.; Chollet, M.; Ravy, S.; et al. Sequential Activation of Molecular Breathing and Bending during Spin-Crossover Photoswitching Revealed by Femtosecond Optical and X-Ray Absorption Spectroscopy. *Phys. Rev. Lett.* **2014**, *113* (22). https://doi.org/10.1103/PhysRevLett.113.227402.
(23) McCusker, J. K.; Walda, K. N.; Dunn, R. C.; Simon, J. D.; Magde, D.; Hendrickson, D. N. Subpicosecond 1MLCT .Fwdarw. 5T2 Intersystem Crossing of Low-Spin Polypyridyl Ferrous Complexes. *J. Am. Chem. Soc.* **1993**, *115* (1), 298–307. https://doi.org/10.1021/ja00054a043.
(24) Consani, C.; Prémont-Schwarz, M.; ElNahhas, A.; Bressler, C.; van Mourik, F.; Cannizzo, A.; Chergui, M. Vibrational Coherences and Relaxation in the High-Spin State of Aqueous [Fe$^{II}$(Bpy)$_3$]$^{2+}$. *Angew. Chem. Int. Ed.* **2009**, *48* (39), 7184–7187. https://doi.org/10.1002/anie.200902728.
(25) Holmes, M. A.; Le Trong, I.; Turley, S.; Sieker, L. C.; Stenkamp, R. E. Structures of Deoxy and Oxy Hemerythrin at 2.0 Å Resolution. *J. Mol. Biol.* **1991**, *218* (3), 583–593. https://doi.org/10.1016/0022-2836(91)90703-9.

(26) Nordlund, P.; Sjöberg, B.-M.; Eklund, H. Three-Dimensional Structure of the Free Radical Protein of Ribonucleotide Reductase. *Nature* **1990**, *345* (6276), 593–598. https://doi.org/10.1038/345593a0.

(27) Rosenzweig, A. C.; Frederick, C. A.; Lippard, S. J.; Nordlund, P. auml;r. Crystal Structure of a Bacterial Non-Haem Iron Hydroxylase That Catalyses the Biological Oxidation of Methane. *Nature* **1993**, *366* (6455), 537–543. https://doi.org/10.1038/366537a0.

(28) Randall, C. R.; Zang, Y.; True, A. E.; Que, L.; Charnock, J. M.; Garner, C. D.; Fujishima, Y.; Schofield, C. J.; Baldwin, J. E. X-Ray Absorption Studies of the Ferrous Active Site of Isopenicillin N Synthase and Related Model Complexes. *Biochemistry (Mosc.)* **1993**, *32* (26), 6664–6673. https://doi.org/10.1021/bi00077a020.

(29) Auböck, G.; Chergui, M. Sub-50-Fs Photoinduced Spin Crossover in [Fe(Bpy)3]2+. *Nat. Chem.* **2015**, *7* (8), 629–633. https://doi.org/10.1038/nchem.2305.

(30) Lemke, H. T.; Kjær, K. S.; Hartsock, R.; van Driel, T. B.; Chollet, M.; Glownia, J. M.; Song, S.; Zhu, D.; Pace, E.; Matar, S. F.; et al. Coherent Structural Trapping through Wave Packet Dispersion during Photoinduced Spin State Switching. *Nat. Commun.* **2017**, *8* (1). https://doi.org/10.1038/ncomms15342.

(31) Terrett, J. A.; Cuthbertson, J. D.; Shurtleff, V. W.; MacMillan, D. W. C. Switching on Elusive Organometallic Mechanisms with Photoredox Catalysis. *Nature* **2015**, *524* (7565), 330–334. https://doi.org/10.1038/nature14875.

(32) Naim, A.; Bouhadja, Y.; Cortijo, M.; Duverger-Nédellec, E.; Flack, H. D.; Freysz, E.; Guionneau, P.; Iazzolino, A.; Ould Hamouda, A.; Rosa, P.; et al. Design and Study of Structural Linear and Nonlinear Optical Properties of Chiral [Fe(Phen)$_3$]$^{2+}$ Complexes. *Inorg. Chem.* **2018**, *57* (23), 14501–14512. https://doi.org/10.1021/acs.inorgchem.8b01089.

(33) Iazzolino, A.; Ould Hamouda, A.; Naïm, A.; Stefánczyk, O.; Rosa, P.; Freysz, E. Nonlinear Optical Properties and Application of a Chiral and Photostimulable Iron(II) Compound. *Appl. Phys. Lett.* **2017**, *110* (16), 161908. https://doi.org/10.1063/1.4981254.

(34) Gallé, G.; Jonusauskas, G.; Tondusson, M.; Mauriac, C.; Letard, J. F.; Freysz, E. Transient Absorption Spectroscopy of the [Fe(2 CH3-Phen)3]2+ Complex: Study of the High Spin↔low Spin Relaxation of an Isolated Iron(II) Complex. *Chem. Phys. Lett.* **2013**, *556*, 82–88. https://doi.org/10.1016/j.cplett.2012.11.034.

(35) Tribollet, J.; Galle, G.; Jonusauskas, G.; Deldicque, D.; Tondusson, M.; Letard, J. F.; Freysz, E. Transient Absorption Spectroscopy of the Iron(II) [Fe(Phen)3]2+ Complex: Study of the Non-Radiative Relaxation of an Isolated Iron(II) Complex. *Chem. Phys. Lett.* **2011**, *513* (1–3), 42–47. https://doi.org/10.1016/j.cplett.2011.07.048.

(36) Tribollet, J.; Galle, G.; Jonusauskas, G.; Deldicque, D.; Tondusson, M.; Letard, J. F.; Freysz, E. Transient Absorption Spectroscopy of the Iron(II) [Fe(Phen)3]2+ Complex: Study of the Non-Radiative Relaxation of an Isolated Iron(II) Complex. *Chem. Phys. Lett.* **2011**, *513* (1–3), 42–47. https://doi.org/10.1016/j.cplett.2011.07.048.

(37) Boyd, R. W. *Nonlinear Optics*, 3rd ed.; Academic Press: Amsterdam ; Boston, 2008.

(38) Kurtz, S. K.; Perry, T. T. A Powder Technique for the Evaluation of Nonlinear Optical Materials. *J. Appl. Phys.* **1968**, *39* (8), 3798–3813. https://doi.org/10.1063/1.1656857.